\documentclass[rnote]{aa} 
\usepackage{graphicx}
\usepackage{txfonts}
\usepackage{natbib}
\begin{document}
   \title{Spectropolarimetric observations of cool DQ white dwarfs}

   \subtitle{}

   \author{T. Vornanen
          \inst{1}
          \and
          S.V. Berdyugina\inst{2,} \inst{3}
          \and
          A. Berdyugin\inst{4}
          }

   \institute{Department of Physics and Astronomy, University of Turku,
              Vaisalantie 20, FI-21500 Piikkio, Finland\\
              \email{tommi.vornanen@utu.fi}
              \and Kiepenheuer-Institut f\"ur Sonnenphysik,
              Sch\"oneckstr.6, D-79104 Freiburg, Germany 
              \and NASA Astrobiology Institute, Institute for Astronomy, University of Hawaii, HI, USA
              \and Finnish Center for Astronomy with ESO, 
               University of Turku, Vaisalantie 20, FI-21500 Piikkio, Finland
             }
   \date{}

 
  \abstract
   {}
   {Following our recent discovery of a new magnetic DQ white dwarf (WD) with CH molecular features, we report the results for the rest of the DQ WDs from our survey.}
   {We use high signal-to-noise spectropolarimetric data to  search for magnetic fields in a sample of 11 objects.}
   {One object in our sample, WD1235+422, shows the signs of continuum circular polarization that is  similar to some peculiar DQs with unidentified molecular absorption bands, but the low S/N and spectral resolution of these data make more observations necessary to reveal the true nature of this object. }
   {}

   \keywords{ magnetic fields - 
              molecular processes -
              white dwarfs
               }

   \maketitle 
%

\section{Introduction}

White dwarfs that have carbon features in any part of the 
electromagnetic spectrum are classified as DQ white dwarfs (WDs).
In these stars, carbon in either atomic or molecular form is the dominant and usually the only 
source of any spectral features. 
The carbon comes from convection induced dredge-up or the atmosphere is largely made up of carbon because of earlier stages of stellar evolution. The latter case is thought to be the origin of hot DQs \citep{duf08}. Cooler members of the group get their carbon from dredge-up. In addition to the hot DQs, there are three other distinct groups of DQ WDs. Cool DQs show molecular absorption bands of C$_2$ and in two cases CH. In peculiar DQs, unidentified molecular features  are broader and deeper, and their spectra look very abnormal. Between the hot DQs that show atomic absorption lines of CII and the cool DQs, there are WDs that show absorption lines of CI. These could be called warm DQs to keep the same naming convention as the other groups.

Along with the mass of the star and chemical composition of its atmosphere, one important element 
of WD physics is magnetism. There have been extensive studies on the magnetic fields of hydrogen WDs 
\citep[see][for example]{kul09}, but the incidence of magnetism in carbon WDs is still unknown. There are 
currently six magnetic cool DQ WDs:  LP790-29 \citep{lie78, wic79, bue99}, LHS2229 \citep{sch99}, 
SDSS J1113+0146 \citep{sch03}, SDSS J1333+0016 \citep{sch03}, G99-37 \citep{ang74}, and GJ841B 
\citep[hereafter Paper I]{vor10}. However, four of these stars belong to the subclass of peculiar DQ WDs with broad, unidentified molecular bands, and only the last two are normal cool DQs.

As is often with negative results, there are very few reports on DQs that have been studied 
and found not to have polarization. \citet{sch99} mention three stars, LHS1126, G225-68, and ESO 439-162, 
for which a search for circular polarization provided no results. However, the authors give upper limits 
of B=3, 2, and 30 MG for the stars, respectively, which certainly do not exclude their magnetic nature. 
To understand magnetism, its origin, and evolution in WDs, it is important that we have information on 
non-magnetic WDs to compare their properties to the properties of their magnetic kin.

As an example of the importance of magnetic fields, we mention the subclass of hot DQs that has now 14 
members. Five of them have been found to be magnetic, and another four might have a substantial magnetic field 
\citep{duf10}. This gives an unusually high rate of magnetism (from 36 \% to 64 \%) for these WDs as compared 
to the average rate of 10-15 \% for all WDs in the solar neighborhood \citep{lie03}. Such a high fraction of magnetic 
objects in a class of stars is unusual. Coupled with the cool, carbon-rich atmosphere, this leads to questions about the role of magnetic fields in their evolution.

In Paper I, we presented our results for GJ841B. In addition to G99-37, GJ841B is the second known DQ WD with circularly polarized spectral 
features of CH-molecules and the first one to apparently have polarized Swan bands of C$_2$. Using the model of molecular polarization in strong magnetic fields by Berdyugina et al. (2007), and appliying it to the CH bands at 430nm and 390nm, 
we determined the field strength of 1.3 MG and the temperature of 6100 K for GJ841B. The discovery of this new magnetic WD came from a survey of 
cool DQs that we have carried out over the past four years. During this survey we have also observed ten other cool DQs 
that do not show circular polarization in optical wavelengths and thus can be considered as non-magnetic. We have 
also discovered circular polarization in WD1235+422, which is a clear indication of its magnetic nature.

\section{Observations}

In our survey of cool DQ WDs, we have observed thirteen DQ WDs using two different instruments at three different times: ALFOSC at the Nordic 
Optical Telescope (NOT) in February 2010, FORS at VLT/UT2 in November 2008, and FORS at VLT/UT1 in August 2009. Details 
of the observations are given in Table~\ref{table1}. In both VLT observing runs, we used Grism 600B+12, which gives wavelength 
coverage from 330 nm to 621 nm with a spectral resolution of $R$=780. This wavelength interval includes several important 
molecular features of C$_2$ and CH. 

Our observations at the NOT were done with low resolution grism \#10, which covers the wavelength region 330-1055 nm. We 
used this grism to get a spectral resolution as low as possible because our targets were faint for a 2.5 m 
telescope. The adopted strategy of a low resolution grism and a wide slit (1.8$\arcsec$) resulted in a spectral 
resolution of only $R$=70. This low resolution gave us a chance to observe our faint targets with spectropolarimetry, 
which normally requires a high S/N ratio as a technique. Our goal was to find polarization in molecular bands that are 
dozens of nanometers wide, so a low resolution was adequate for our purposes.

\begin{figure}
\resizebox{8.8cm}{!}{\includegraphics{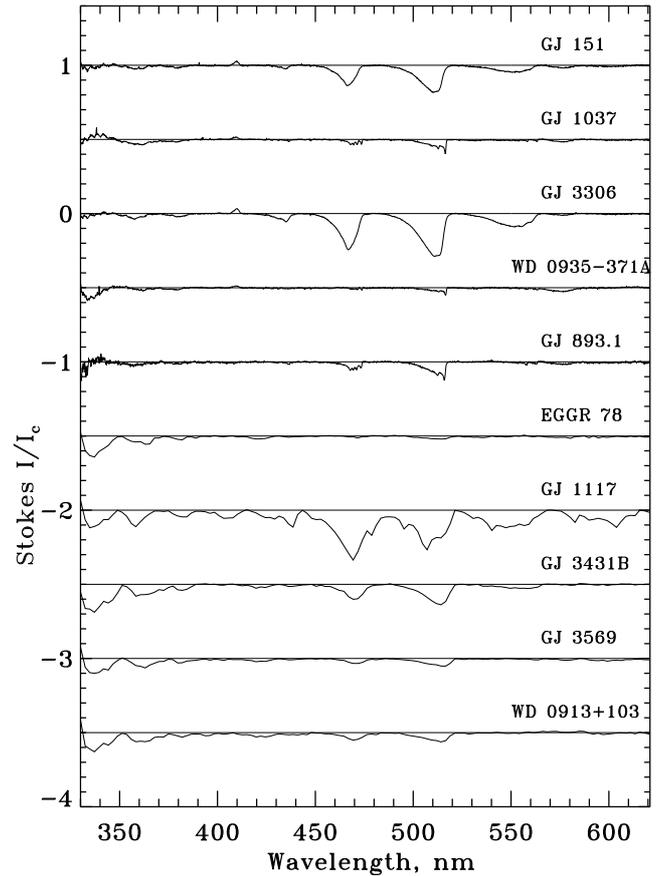}}
\caption{Continuum normalized intensity spectra of the observed stars. Spectrum of GJ151 is at its actual value and the rest are shifted in intervals of 0.5. Horizontal lines denote the continuum level of each spectrum. See Fig. \ref{fig:WD1235I} for WD1235+422. } 
\label{fig:intensity}
\end{figure}

\begin{figure}
\resizebox{8.8cm}{!}{\includegraphics{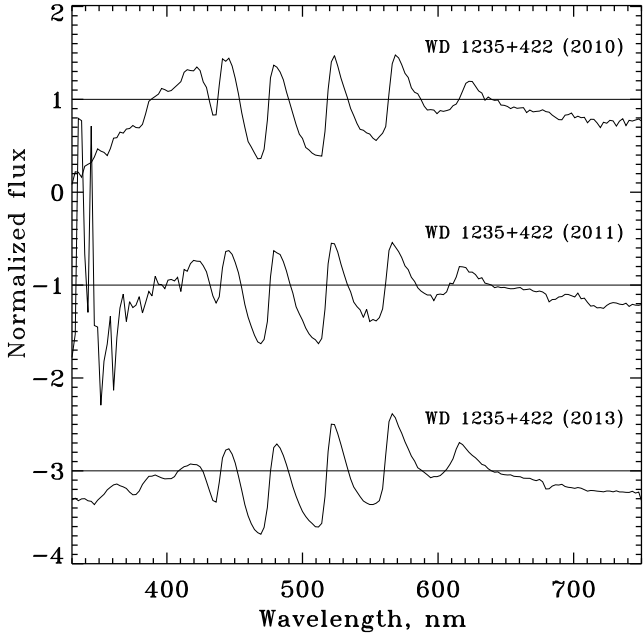}}
\caption{The spectra of WD1235+422 in three epochs are normalized to the flux value at 639.4 nm, and the horizontal lines denote this value. The bottom two spectra are shifted vertically.}  
\label{fig:WD1235I}
\end{figure}

In the data reduction process, we subtracted the bias level from all images and also removed cosmic ray contamination. We skipped the flatfielding, because it did not improve the quality of data and sometimes even made it worse. Background was subtracted separately for the two beams, and the spectra were extracted. Wavelength calibration was done by using just one arclamp spectrum and not by using lamp spectra in different retarder plate positions. Finally, a dispersion correction was applied using the number of pixels in the dispersion direction of the raw image as the number of wavelength points. Spectra were normalized to the continuum level, because the model also produces normalized spectra. The intensity spectra of our targets are shown in Figs.~\ref{fig:intensity} and ~\ref{fig:WD1235I}. 

\begin{figure}
\resizebox{8.8cm}{!}{\includegraphics{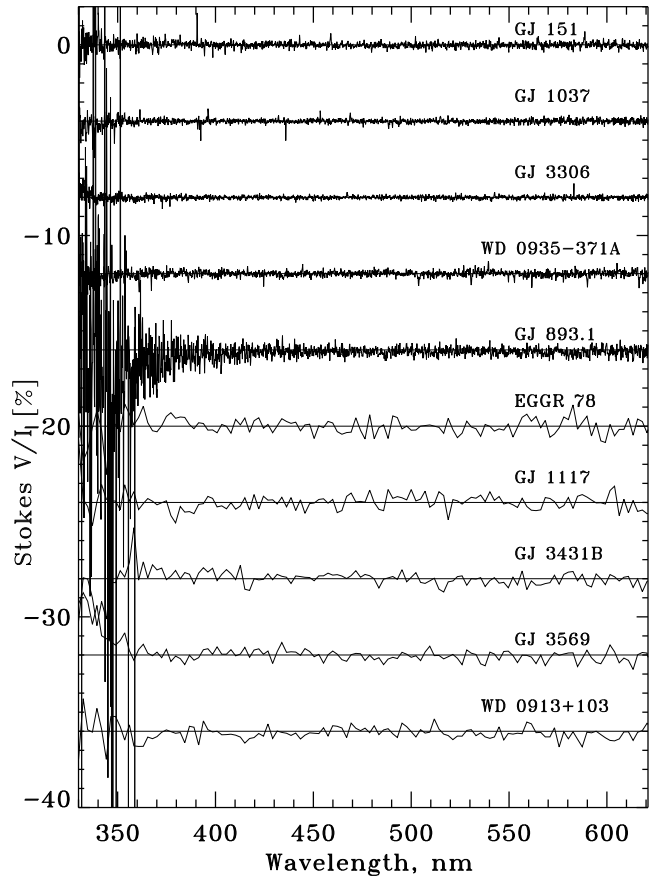}}
\caption{Stokes V/I spectra of the observed stars (in percents). Spectrum of GJ151 is at its real value, and the rest of the spectra are shifted in intervals of 4 \%. See Fig. \ref{fig:WD1235V} for WD1235+422.} 
\label{fig:stokesv}
\end{figure}

In our survey, we have decided to concentrate on circular polarization (Stokes $V$). From previous studies on G99-37 \citep{ber07}, we 
expect a larger signal in circular than in linear polarization. Circular polarization observations require fewer measurements and calibrations. This allows us to dedicate more time for actual scientific exposures and reduce overheads. In observations made with VLT, our goal was to achieve a polarization accuracy of $\sigma_P=\pm 
0.2 \%$ (S/N$\sim$700). In the case of NOT observations, we could not get such a high accuracy, but nevertheless managed 
to attain approximately S/N=300 ($\sigma_P=\pm 0.5 \%$). The polarization spectra are shown in Fig.~\ref{fig:stokesv}. 

\begin{figure}
\resizebox{8.8cm}{!}{\includegraphics{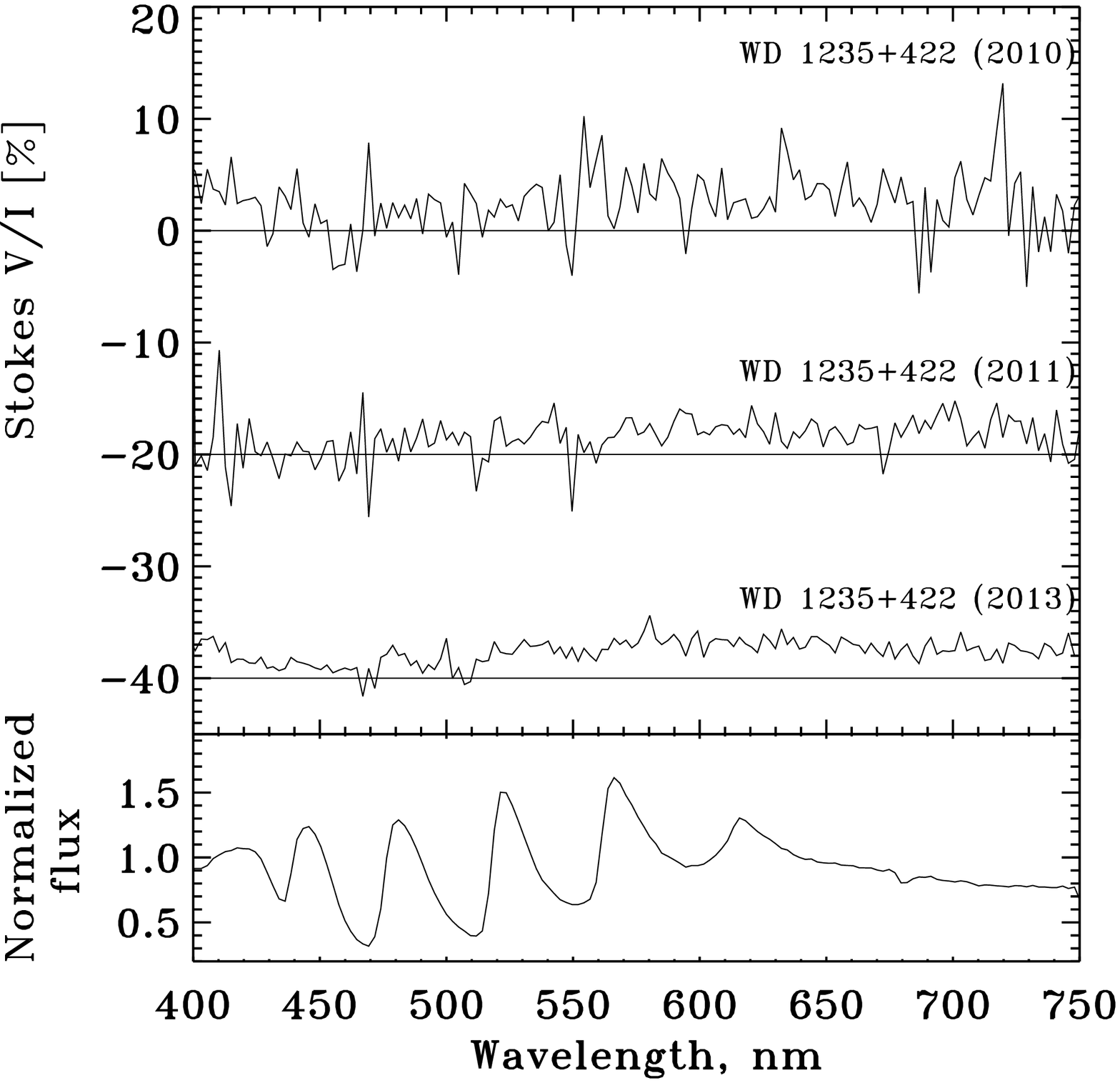}}
\caption{Stokes V/I spectra of WD125+422 from 2010, 2011, and 2013. Zero levels are marked with horizontal lines and the spectra have been displaced for clarity. Intensity spectrum in the bottom panel is for the comparison of feature wavelengths and clearly indicates that the polarization is of continuum origin. } 
\label{fig:WD1235V}
\end{figure}

Additional observations of one of our targets, WD1235+422, were taken in service mode with the NOT on May 21st, 
2011 and again on March 20th, 2013. The S/N of the 2011 measurement was better than in our 2010 observations (see Figures \ref{fig:WD1235I} and \ref{fig:WD1235V}) due to the doubled exposure time with the 2013 measurement almost doubling the exposure time again. The 
resulting average spectrum is similar in all observing runs, but the poor spectral resolution does not allow us to make any conclusions about the magnetic properties, like field strength or geometry, of WD1235+422, apart from the conclusion that the star is indeed magnetic.

The intensity spectra in Fig.~\ref{fig:intensity} are normalized to continuum, but the absorption bands in WD1235+422 are overlapping, so there is no continuum to fit between 430 nm and 630 nm. For this reason we have flux calibrated the spectra and then normalized them to the value at 639.4 nm. In this way, some comparison to possible future observations and modelling can be done.

We note that the little emission peak at 410\,nm in the objects observed with FORS1 (the first four in Figures \ref{fig:intensity} and \ref{fig:stokesv}) is not intrinsic to the objects. It is a spurious reflection signal in the optics, since it is present in the 
raw images of every object as a bright spot very close to the spectra.

\begin{table*}
\caption{Observing log for our survey of cool DQ white dwarfs.}
\label{table1}
\centering
\begin{tabular}{lllccccc}
\hline\hline
Object & RA	& Dec	& M$_V$ & Observing Date & Exp Time (s) & R=$\lambda$/$\vartriangle\lambda$ & Instrument \\
\hline

\object{GJ1037} & 01 18 00.086 & +16 10 20.63 & 13.85 & 15.11.2008 & 3300 & 780 & VLT/FORS1 \\
\object{GJ151} & 03 44 34.860 & +18 26 09.77 & 15.2 & 16.11.2008 & 9600 & 780 & VLT/FORS1 \\
\object{GJ3306} & 04 37 47.42 & -08 49 10.7 & 14.1 & 16.11.2008 & 4800 & 780 & VLT/FORS1 \\
\object{WD0935-371A} & 09 37 00.0 & -37 21 33 & 14.85 & 15.11.2008 & 7200 & 780 & VLT/FORS1 \\
\object{GJ893.1} & 23 14 25.191 & -06 32 47.84 & 15.42 & 26.-27.8.2009 & 9000 & 780 & VLT/FORS \\
\object{GJ3431B} & 07 10 14.332 & +37 40 19.51 & 15.94 & 10.2.2010 & 8400 & 70 & NOT/ALFOSC \\
\object{GJ1117} & 08 59 14.714 & +32 57 12.11 & 15.18 & 11.2.2010 & 3600 & 70 & NOT/ALFOSC \\
\object{WD0913+103} & 09 16 02.69 & +10 11 10.7 & 15.77 & 10.-11.02.2010 & 9600 & 70 & NOT/ALFOSC \\
\object{GJ3569} & 09 50 17.139 & +53 15 14.95 & 15.3 & 10.-11.02.2010 & 7200 & 70 & NOT/ALFOSC \\
\object{EGGR78} & 11 18 15.12 & -03 14 05.4 & 15.39 & 10.2.2010 & 3600 & 70 & NOT/ALFOSC \\
\object{WD1235+422} & 12 37 52.240 & +41 56 24.79 & 16.8 & 10.-11.2.2010 & 3600 & 70 & NOT/ALFOSC \\
 &  &  &  & 21.5.2011 & 7200 & 70 & NOT/ALFOSC \\
 &  &  &  & 20.3.2011 & 12000 & 70 & NOT/ALFOSC \\
\hline
\end{tabular}
\end{table*}

\section{Results}
\label{sec:results}

The intensity spectra of the objects in Table \ref{table1} are shown in Figure \ref{fig:intensity} and Stokes V/I in 
Figure \ref{fig:stokesv}. Our sample of DQ WDs shows a large variation in the depth of C$_2$ features from 3 \% to 30 \%
below the continuum level. The shape of the absorption features can be divided into two distinct categories. For example, see the $\Delta v=0$ Swan band of GJ1037 at $\lambda=510$ nm and compare it with the same band of GJ3306, where it is rounded and shows no internal structure. At high temperatures ($T \geq 7000$K), the higher energy vibrational bands with $v\geq3$ are visible as the individual dips within the Swan bands. We would like to stress that this difference in appearance is real. All stars observed with VLT have a spectral resolution of about 0.7 nm.

For the objects observed with the NOT, the spectral resolution ranges from 5 nm to 15 nm across the observed wavelength region, which is too poor to reveal these features, but it is worth to 
note that one of these objects, GJ1117 (WD0856+331), shows lines of neutral carbon. The lines look very broad, but this 
is just an effect of the low resolution \citep[See Figure 7 from][for a higher resolution spectrum]{duf05}. These 
lines disappear at temperatures below 9000 K, so they provide an immediate temperature estimate for this star. The 
simultaneous existence of C$_2$ gives an upper limit of 11000 K.

The Stokes V spectra of the survey targets, apart from GJ841B, show no signatures of polarization with one exception: WD1235+422 (See Figure \ref{fig:WD1235V}) shows a large circular polarization signal of a few percent across the observed wavelength region in all sets of observations taken with the NOT over four years. It also has the strongest absorption bands of C$_2$ among all stars presented in this paper. In addition, its molecular features are slightly shifted from the normal positions of Swan bands. This would make WD1235+422 related to the class of peculiar DQ WDs \citep{sch03,hal08}. The Stokes V/I spectrum does not show any significant variations in the Swan bands at this resolution except for a slight decrease in the degree of polarization. Similar behaviour is shown by LP 790-29, one of the magnetic peculiar DQs \citep{sch99}. Thus, we conclude that the 2-3\%\ circular polarization in WD1235+422 is due to continuum polarization.

The white dwarf WD1235+422 seems to be an interesting link between normal and peculiar DQ WDs and thus deserves a more careful examination. Interestingly, the intensity spectrum of WD1235+422 looks similar to the simulated spectrum of a 10000 K purely carbon atmosphere DQ WD, as seen in Fig. 6 of \citet{duf08}. On the other hand, \citet{koe06} modelled it as peculiar DQ WD with a temperature of 5846 K based on photometry. There are at least two other DQ WDs with similar spectra: \object{GSC2U J131147.2+292348} \citep{car03} and SDSS J090208.40+20104 \citep{lim13}.  
We think that some attention needs to be given to these stars in the future because they might give new insight into the evolution of hot DQs.

We have tried to model the observed spectra of these WDs using the method by \citet{ber07}. This model is based on the Unno-Rachkovsky and Milne-Eddington approximations and is capable of inferring the temperature and magnetic field strength in a single-layer atmosphere.  Using this model, we have encountered a problem to fit the C$_2$ band strengths simultaneously with a single set of the parameters. We assume that this difficulty is related to the single-layer approximation and can be overcome with more realistic, stratified atmospheres of WDs, such as that by \citet{duf05}. We consider these atmospheres and address the modelling problems in a separate, forthcoming paper.

\section{Conclusions}

We have presented low resolution circular spectropolarimetric observations of 11 DQ WDs. Among these stars we found a 
circular polarization signal in only one, WD1235+422, with the addition of the previously reported GJ841B. 

Although we did not find any other magnetic WDs in our sample apart from GJ841B and WD1235+422, the data are still valuable for the 
study of DQ WDs. At the moment, there are about 182 confirmed cool DQ WDs, and by including WD1235+422, only seven of them have been
found to be magnetic. (There are also fourteen hot DQs of which at least five are magnetic.) This amounts to 4 \% of the whole population, and the value is exactly what has been found for DA white dwarfs in the Sloan Digital Sky Survey (SDSS) \citep{kep13}. The study of \citet{kep13} only concerns magnetic field strengths of over 2 MG, and these are also the values found for cool DQs so far. The authors note that there should be about as many low field WDs than these higher field dwarfs, but they are very difficult to find. \citet{lan12} report results on a field of 10 kG on a DA white dwarf.

Magnetism seems to be a good way to study the connections between the different subclasses of DQ WDs. However, no one to our knowledge has made an extensive study of the magnetism of those DQ WDs that have absorption lines of neutral carbon. These might be called warm DQs, because they are between the hot DQs (CII lines) and cool DQs (molecular absorption) in temperature. One such object has been found recently \citep{wil13}. However, a lot still remains to be done, but we are hopefully moving in the right direction in solving the mystery of carbon atmosphere white dwarfs.

\begin{acknowledgements}
This work was supported by the Leibniz Association grant SAW-2011-KIS-7.
\end{acknowledgements}

\bibliographystyle{aa}

\bibliography{/home/ttvorn/Desktop/omat_paperit/References/bibliography}



\end{document}